\documentclass[aps,prb,superscriptaddress,floats,reprint,nofootinbib]{revtex4-2}
\usepackage[colorlinks=true,
    linkcolor=blue,
    urlcolor=blue,
    citecolor=green]{hyperref}

\usepackage[english]{babel}
\usepackage[utf8]{inputenc}
\usepackage{amsfonts,graphicx,soul,mathrsfs}
\usepackage{url}
\usepackage[dvipsnames]{xcolor}
\usepackage[normalem]{ulem}
\usepackage{epstopdf}
\usepackage{textcomp}
\usepackage{xcolor}
\usepackage{soul}
\usepackage{physics}
\usepackage{graphics}
\usepackage{comment}
\usepackage{amssymb}
\usepackage{amsmath}
\usepackage{tensor}

\usepackage{subcaption}

\usepackage{animate}

\newcommand{\be}{\begin{equation}}
\newcommand{\ee}{\end{equation}}
\newcommand{\bear}{\be\begin{array}}
\newcommand{\bw}{\begin{widetext}}
\newcommand{\ew}{\end{widetext}}
\newcommand{\bea}{\begin{eqnarray}}
\newcommand{\eea}{\end{eqnarray}}

\usepackage{xcolor}

\usepackage{dcolumn}
\usepackage{bm}

\begin{document}

\title{Material criteria for angular-momentum transfer from twisted photons to photoelectrons}
\author{A. A.~Shchepkin}
\email{alexander.shchepkin@gmail.com}
\affiliation{{School of Physics and Engineering, ITMO University, 197101 St.\,Petersburg, Russia}}
\affiliation{{Petersburg Nuclear Physics Institute named by B. P. Konstantinov of National Research Centre ``Kurchatov Institute,'' 188300 Gatchina, Russia}}

\author{D. V.~Karlovets}
\affiliation{{School of Physics and Engineering, ITMO University, 197101 St.\,Petersburg, Russia}}
\affiliation{{Petersburg Nuclear Physics Institute named by B. P. Konstantinov of National Research Centre ``Kurchatov Institute,'' 188300 Gatchina, Russia}}

\begin{abstract}
We study the material requirements for transferring the total-angular-momentum (TAM) projection of a twisted photon to the orbital angular momentum (OAM) of a photoelectron emitted from an $n$-type semiconductor surface. Within a single-valley effective-mass model of surface photoemission, we calculate OAM-resolved spectra for GaAs, InSb, $\mathrm{InAs}_{0.34}\mathrm{Sb}_{0.66}$, and $\mathrm{Hg}_{0.76}\mathrm{Cd}_{0.24}\mathrm{Te}$ irradiated { by twisted photons with Laguerre--Gaussian--like momentum profile}. The OAM transfer is characterized by the mean electron OAM, its root-mean-squared width, and { the fraction of electrons with OAM-projection equal to the photon TAM-projection}. Efficient transfer requires the transverse momentum of the conduction electron to be much smaller than the transverse momentum width of the photon, which favors low temperatures, small conduction-band effective masses, and shallow donor levels. %At temperature $T=0.5$~K and { photon TAM projection $m_\gamma=1$}, the desired-OAM fraction reaches $87$--$91\%$ for InSb, $\mathrm{InAs}_{0.34}\mathrm{Sb}_{0.66}$, and $\mathrm{Hg}_{0.76}\mathrm{Cd}_{0.24}\mathrm{Te}$. For larger photon TAM projections, however, the distributions broaden substantially: at $m_\gamma=64$, only $16$--$17\%$ of the electrons occupy the desired OAM channel even though the mean OAM remains close to $m_\gamma$. Raising the temperature to $4$~K further reduces the selectivity. 
GaAs is less favorable within the present model because of its larger effective mass and deeper donor levels. 
%although its conventional negative electron affinity operating regime lies outside this model. 
We also discuss the demanding photon-wave packet parameters, possible extensions to Dirac and Weyl semimetals, and post-emission OAM selection. 
%In particular, a log-polar sorter supplemented by a fan-out transformation and spatial postselection can isolate the desired OAM component without requiring mutual coherence between different OAM channels. 
Although the generation of vortex electrons with a vanishing OAM uncertainty via photoemission seems challenging, we argue that it is still possible to generate single-electron vortex states with a finite OAM dispersion.

\end{abstract}

\maketitle

\section{Introduction}

Structured states of light and matter have become important tools in quantum optics, electron microscopy, condensed-matter physics, and accelerator science. Vortex, or twisted, states are distinguished by a phase singularity and a well-defined projection of angular momentum onto a selected propagation axis \cite{Allen1992,Uchida2010,Verbeeck2010}. For a free electron, a state with an azimuthal phase factor $\exp(il\varphi)$ carries an orbital-angular-momentum (OAM) projection $L_z=\hbar l$. This additional degree of freedom can be used to probe magnetic and chiral properties of matter, characterize spatial symmetries, and introduce new observables in scattering and radiation processes \cite{Verbeeck2010,Juchtmans2015}. {Broader reviews and recent developments of vortex-particle physics can be found in \cite{revBliokh2017,revIvanov2022,revKarlovetsSerbo2020,revKarlovets2022EPJC,revKarlovets2023EPJC}.}

Several methods for producing vortex electrons have already been demonstrated. These include nanofabricated amplitude and phase holograms \cite{Uchida2010,Verbeeck2010,McMorran2011}, material and magnetic phase plates \cite{Beche2014}, aberration-controlled electron lenses \cite{Clark2013}, and astigmatic mode converters \cite{Schattschneider2012,Kramberger2019,Schachinger2021}. Although these methods have enabled extensive studies in transmission electron microscopy, none of them simultaneously provides arbitrary OAM, high current, high transmission, ultrashort pulse duration, and straightforward compatibility with an accelerator photoinjector. The development of an intense and scalable source of vortex electrons therefore remains an open experimental problem.

{ Photoemission driven by twisted light provides an attractive alternative, as also suggested by the demonstrated transfer of the photon total-angular-momentum (TAM) projection to the electron OAM in atomic photoionization \cite{revPavlov2024}.} If the azimuthal phase structure of the electromagnetic field is transferred to an emitted electron, the photocathode can act as the first stage of a vortex-electron source. Such a source could generate short electron pulses synchronized with the drive laser and subsequently accelerate them to relativistic energies. Unlike intercepting electron-optical elements placed after acceleration, the transfer would occur directly at the cathode surface \cite{Kazinski2025}.

An accelerator implementation requires structured radiation in the deep-ultraviolet (DUV) range, because the photon energy must exceed the relevant photoemission threshold. Recent experiments at the Joint Institute for Nuclear Research (JINR) in Dubna, Russia, have demonstrated the generation of optical vortices at wavelengths of approximately 260--266 nm using amplitude and phase spiral zone plates \cite{Dyatlov2025}. A phase spiral zone plate produced a beam with the topological charge $l_\gamma=1$ and a conversion efficiency of approximately 40\% \cite{Dyatlov2025}. More recently, fork gratings, a spiral phase plate, and binary axicons were integrated into an RF-photoinjector laser line. The spiral phase plate generated a high-purity mode with $l_\gamma=64$ and a measured optical conversion efficiency of approximately 80\% \cite{Dyatlov2026}. These developments establish experimentally relevant DUV wavelengths and accessible topological charges, although the tightly focused broadband packets used in the calculations remain more demanding than the demonstrated optical states.

A well-defined angular momentum of the incident photon does not, however, guarantee that the emitted electron occupies a single OAM eigenstate. { Although perfect transfer of the photon TAM projection to the OAM of the emitted electron is possible for photoionization of a single atom placed on the photon propagation axis \cite{revPavlov2024}, the initial electron in a solid has a finite momentum distribution determined by the band structure, temperature, carrier concentration, and doping.} Surface roughness and defects can additionally break the axial symmetry required for OAM conservation. Consequently, even a nearly pure incident photon mode can generate a photoelectron distribution extending over several values of \(l_e\) \cite{Kazinski2025}.

{ The recent work by Kazinski et al.~\cite{Kazinski2025} has developed a theory of the surface photoelectric effect induced by twisted photons and has derived the probability of detecting an emitted electron with specified longitudinal momentum, transverse-momentum magnitude, and OAM projection. Their analysis shows that the leading contribution obeys the approximate selection rule}

$$ l_e=m_\gamma, $$
where \(m_\gamma\) is the total angular-momentum projection of the photon state. Corrections are controlled by the transverse momentum of the initial electron. Nearly selective transfer requires

$$ p_{i,\perp}\ll p_{f,\perp}\sim k_\perp, $$
where $p_{i,\perp}$, $p_{f,\perp}$ and $k_\perp$ are the transverse momentum components of the conduction electron, of the photoelectron state out of the cathode and of the incident photon, respectively. Lightly doped n-InSb was identified as a favorable material because of its small conduction-band effective mass and shallow donor levels. Dirac and Weyl semimetals were also suggested as prospective candidates owing to their linear low-energy dispersion \cite{Kazinski2025}.

The material with the smallest effective mass is not necessarily the most suitable experimental photocathode. A realistic assessment must also account for the density of conduction states, donor freeze-out, surface preparation, electron affinity or crystal-vacuum energy offset, chemical stability, and the optical response at the incident wavelength. In particular, increasing the doping can provide more conduction electrons but also increases the Fermi momentum and may broaden the OAM distribution. Conversely, extremely light doping can improve OAM selectivity while reducing the emitted charge.

{ In this work, we apply the theory from the work \cite{Kazinski2025} to a wider range of photocathode materials and determine how the effective mass, donor binding energy, carrier concentration, and temperature control OAM transfer. InSb is compared with the narrow-gap alloys \(\mathrm{InAs}_{0.34}\mathrm{Sb}_{0.66}\), \(\mathrm{Hg}_{0.76}\mathrm{Cd}_{0.24}\mathrm{Te}\), and with GaAs under a common set of photon-wave-packet parameters. The results demonstrate that the mean emitted OAM alone can substantially overestimate the selectivity of the transfer and should be considered together with the fraction in the desired OAM state and the spectral width.} 

For the three narrow-gap materials (InSb, \(\mathrm{InAs}_{0.34}\mathrm{Sb}_{0.66}\) and \(\mathrm{Hg}_{0.76}\mathrm{Cd}_{0.24}\mathrm{Te}\)), we calculate the OAM-resolved spectrum of the emitted electrons and characterize it by the mean OAM, { the fraction of electrons with OAM equal to the angular momentum of the photon, further referred to as the \textit{desired-OAM fraction}}, and the root-mean-squared (rms) spectral width. This allows us to examine whether the residual OAM mixture can be analyzed and conditionally purified using electron OAM sorters. Particular attention is paid to log-polar sorting supplemented by a fan-out transformation, which can improve the separation of neighboring OAM channels without requiring coherence between different OAM components \cite{Mirhosseini2013,Ruffato2018}.

Natural units $\hbar=c=k_{\mathrm B}=1$ are used throughout the paper.

\section{Theory of the surface photoelectric effect induced by twisted photons}
\label{sec:surface-photoelectric-effect}

\begin{figure*}[t]
\centering
\includegraphics[width=1\linewidth]{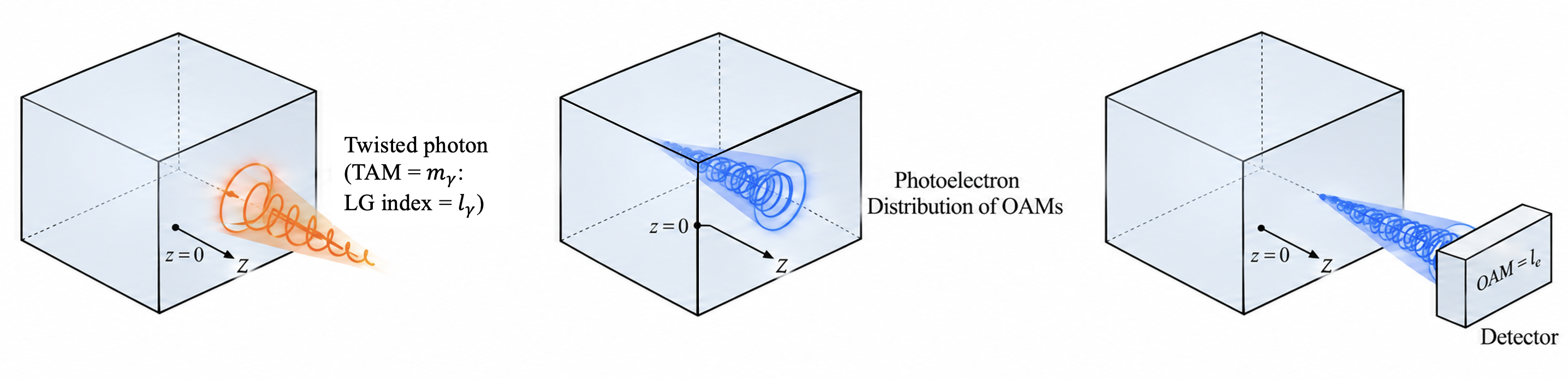}
\caption{Geometry of the surface photoelectric effect induced by a twisted photon. The semiconductor occupies the half-space \(z<0\), and the emitted electron propagates into the vacuum region \(z>0\).}
\label{fig:illust}
\end{figure*}

We employ the theory of surface photoemission induced by twisted photons developed in \cite{Kazinski2025}. The semiconductor occupies the half-space \(z<0\), while \(z>0\) is vacuum. The \(z\) axis is normal to the crystal--vacuum interface and defines the quantization axis for the angular momentum, as illustrated in Fig.~\ref{fig:illust}.

The interface is assumed to be flat, laterally homogeneous, and free of defects. Conduction electrons are described in the single-valley effective-mass approximation, with an approximately parabolic dispersion near the conduction-band minimum. Their spin is not treated explicitly and enters only through the twofold degeneracy of the electronic states. The single-electron Hamiltonian is written as
\begin{equation}
\hat h_0
=
\hat{\bm p}
\frac{1}{2\widetilde M(z)}
\hat{\bm p}
+U(z).
\label{eq:single-particle-hamiltonian}
\end{equation}
{Here \(\hat{\bm p}=-i\bm{\nabla}\) is the electron momentum operator in natural units, \(M_*\) is the conduction-band effective mass, and \(M_0\) is the free-electron mass. The position-dependent mass is \(\widetilde M(z)=M_*\) inside the semiconductor and \(\widetilde M(z)=M_0\) in vacuum.} The potential \(U(z)\) includes the offset
\(-U_0\) of the conduction-band minimum relative to the vacuum level, the
image-charge interaction, and Debye screening inside the semiconductor. The
microscopic profile of the potential across the transition layer is not
required explicitly because it only weakly affects the angular dependence of
near-threshold photoemission.

The initial conduction-electron states are thermally populated according to
the Fermi--Dirac distribution,
\begin{equation}
f_i
=
\frac{1}
{\exp[(E_i-\mu)/T]+1},
\label{eq:fermi-dirac-density-matrix}
\end{equation}
where \(E_i\) is the initial-electron energy measured from the vacuum level
and \(\mu\) is the equilibrium chemical potential. {The symbol \(T\) denotes the photocathode temperature in energy units, because the Boltzmann constant \(k_{\mathrm B}=1\).} Thermionic emission is
neglected, so only initially occupied states with \(E_i<0\) are included.
Electron--electron, electron--phonon, and electron--impurity interactions are
neglected in the photoemission amplitude. Doping enters the calculation through the donor concentration and ionization energy, which determine the equilibrium chemical potential, free-carrier density, screening length, and momentum distribution of the conduction electrons.

{The detected photoelectron is projected onto a Bessel state specified by the OAM projection \(l_e\), the longitudinal momentum projection, \(p_{f,z}\), and the transverse momentum projection, \(p_{f,\perp}\),}
\begin{equation}
\langle\bm r|l_e,p_{f,z},p_{f,\perp}\rangle
=
\frac{J_{l_e}(p_{f,\perp}\rho)}
{\sqrt{2RLp_{f,\perp}}}
\exp\left[i\left(l_e\varphi+p_{f,z}z\right)\right],
\label{eq:twisted-electron-state}
\end{equation}
where \(\bm{r} = (\rho,\varphi,z)\) - coordinate vector in cylindrical representation. {Here, \(l_e\) is the electron OAM projection onto the surface normal, \(J_{l_e}\) is the Bessel function of the first kind, while \(R\) and \(L\) are the radius and length of the cylindrical quantization volume used for normalization.}

The momentum-space amplitude of the incident photon state is taken in the form
{
\begin{equation}
d_\lambda(\bm k)
=
\delta_{\lambda \lambda_\gamma}
g_{nl_\gamma}(k_\perp,k_z)
e^{im_\gamma\varphi_k},
\label{eq:coherent-photon-amplitude}
\end{equation}
}
{where $k_\perp$ and $k_z$ are the transverse and longitudinal photon momentum projections, \(\lambda_\gamma=\pm1\) is the photon helicity and \(m_\gamma\) is the projection of its TAM onto the \(z\) axis. The Kronecker delta \(\delta_{\lambda\lambda_\gamma}\) restricts the plane-wave components of the photon wave packet to a definite helicity \(\lambda = \lambda_\gamma\). In the paraxial limit, the optical topological charge \(l_\gamma\) is related to these quantum numbers by}
{\begin{equation}
 m_\gamma\simeq l_\gamma+\lambda_\gamma.
\end{equation}}
We retain \(m_\gamma\) because it is the exact TAM entering the
photoemission selection rule. Outside the paraxial limit it should not be
identified with the orbital part alone. The longitudinal momentum is
negative for a photon propagating towards the surface,
\begin{equation}
k_z
=
-\sqrt{\omega^2-k_\perp^2},
\label{eq:dispersion_law}
\end{equation}
where $\omega$ is the photon energy. Thus, the exact photon dispersion relation is retained and no paraxial approximation \(\omega\simeq |k_z|\) is imposed.

We use a Laguerre Gaussian (LG) type momentum-space envelope,
\bw
\bea
g_{nl_\gamma}(k_\perp,k_z)
=
\frac{1}
{\sqrt{(2\pi)^{3/2}\sigma_z\sigma_\perp^2}}
\sqrt{\frac{n!}{(n+|l_\gamma|)!}}
\left(
\frac{k_\perp}
{\sqrt{2}\sigma_\perp}
\right)^{|l_\gamma|}
L_n^{|l_\gamma|}
\left(
\frac{(k_\perp)^2}
{2\sigma_\perp^2}
\right)
\exp\left[
-\frac{(k_\perp)^2}{4\sigma_\perp^2}
-\frac{(k_z-\bar k_z)^2}{4\sigma_z^2}
\right],
\label{eq:LG-momentum-4.6}
\eea
\ew
normalized by
\begin{equation}
\int d^3\bm k
\left|
g_{nl_\gamma}(k_\perp,k_z)
\right|^2
=1.
\end{equation}
{This is the standard three-dimensional momentum-space normalization in the laboratory frame, however, it is not a Lorentz-invariant normalization.}
Here, $L_n^{|l_\gamma|}(x)$ is the generalized Laguerre polynomial, \(n\) is the radial quantum number, \(\bar k_z<0\) is the central
longitudinal momentum, and \(\sigma_\perp\) and \(\sigma_z\) are the transverse
and longitudinal momentum widths, respectively. Because the exact dispersion relation is
retained, expression \eqref{eq:LG-momentum-4.6} should be understood as an LG-type
momentum-space wavepacket rather than as a paraxial optical Laguerre--Gaussian beam. In the present calculations only the lowest radial photon mode $n = 0$ is studied.

To the leading nonvanishing order in the electromagnetic interaction, and after
neglecting thermionic emission, {the probability of detecting a photoelectron with OAM projection \(l_e\), longitudinal momentum \(p_{f,z}\), and transverse-momentum magnitude \(p_{f,\perp}\), per unit longitudinal and transverse momentum, is}

\be
\begin{aligned}
& \frac{dP_{l_e}}
{dp_{f,z} dp_{f,\perp}}
=
\frac{e^2}{(2\pi)^4}
\int d^2\bm p_{i,\perp} \\
& \times\int_0^\infty dp_{i,z}\,
\theta(-E_i)
\frac{p_{f,\perp}}{\Delta E_{fi}}
\frac{
\left|\mathcal M_{fi}\right|^2
}{
\exp[(E_i-\mu)/T]+1
}.
\label{eq:differential-probability}
\end{aligned}
\ee

Here, $e$ is the elementary charge, $\theta(x)$ is the Heaviside step function, \(\bm p_{i,\perp}\) is the conduction electron transverse-momentum vector,
\(p_{i,\perp}=|\bm p_{i,\perp}|\) is its magnitude, and \(p_{i,z}\) is the conduction electron longitudinal-momentum component.
\bea
    E_i
    =
    \frac{(p_{i,\perp})^2+(p_{i,z})^2}{2M_*}
    -U_0,
\eea
and
\bea
    \Delta E_{fi}=E_f-E_i,
    \qquad
    E_f
    =
    \frac{(p_{f,\perp})^2+(p_{f,z})^2}{2M_0}.
\eea
The amplitude \(\mathcal M_{fi}\) {depends on} the photon wavepacket, the electron wave functions at the interface, the dielectric response of the material, and the energy- and transverse-momentum-conservation conditions. Its explicit form is given in \cite{Kazinski2025} and is evaluated numerically in the present calculations. In the present calculations, the photoemission probability is averaged over the two photon-helicity states, $\lambda_\gamma=\pm1$.

For an initial electron with zero transverse momentum, axial symmetry gives
the exact selection rule \(l_e=m_\gamma\). {This is analogous to the exact on-axis OAM transfer found for single-atom photoionization by a twisted photon \cite{revPavlov2024}.} At small but finite
\(p_{i,\perp}\), the amplitude has the schematic expansion
\begin{equation}
\mathcal M_{fi}
=
\mathcal M_{fi}^{(0)}
\delta_{l_e,m_\gamma}
+
\frac{p_{i,\perp}}{p_{f,\perp}}\,
\mathcal M_{fi}^{(1)}(\delta_{l_e,m_\gamma+1} - \delta_{l_e,m_\gamma-1})
+\cdots,
\label{eq:oam-amplitude-expansion}
\end{equation}
where \(\mathcal{M}_{fi}^{(0)}\) and \(\mathcal{M}_{fi}^{(1)}\) denote the leading and first-order terms of the transition amplitude, respectively. The first correction populates the neighboring channels \(l_e=m_\gamma\pm1\), while higher orders produce larger deviations from \(l_e=m_\gamma\). Selective OAM transfer therefore requires the characteristic transverse momentum of the initial electrons to satisfy
\begin{equation}
p_{i,\perp}\ll p_{f,\perp},
\qquad
p_{i,\perp}\ll\sigma_\perp.
\label{eq:material-oam-condition}
\end{equation}
{From here, taking into account the dispersion relation \eqref{eq:dispersion_law}, one can also obtain the condition for the longitudinal momentum width of the photon:
\begin{equation}
\delta k_z = \frac{\partial k_z}{\partial k_\perp}\delta k_\perp = \frac{k_\perp}{k_z} p_{i,\perp}\ll \sigma_z,
\end{equation}

Here, $\delta k_\perp$ and $\delta k_z$ do not represent a physical change of the incident photon momentum caused by the initial electron. Rather, for a fixed detected photoelectron momentum, the initial-electron momentum can change to the value depending on which momentum component of the incident photon wave packet satisfies the conservation laws and contributes to the transition. In particular, a transverse variation of the initial-electron momentum corresponds to 
\[
|\delta k_\perp|\sim p_{i,\perp},
\] 
while the photon dispersion relation \eqref{eq:dispersion_law} induces the associated longitudinal shift $\delta k_z$. Requiring $|\delta k_z|\ll\sigma_z$ ensures that these kinematically selected photon components remain within the same narrow region of the longitudinal wave-packet envelope, so that its amplitude is only weakly affected by the initial-electron momentum spread.} These inequalities connect the OAM selectivity directly to the photon wavepacket and to the effective mass, temperature, carrier concentration, and chemical potential of the
photocathode.

\section{Results}
\label{sec:results}

\subsection{Calculation parameters}

We consider four representative $n$-type semiconductors: GaAs, InSb, InAs$_{0.34}$Sb$_{0.66}$, and Hg$_{0.76}$Cd$_{0.24}$Te. These materials are selected because they combine practical availability with favorable electronic properties relevant to OAM transfer. GaAs is included as a technologically mature photocathode material widely employed in electron sources, providing a practical reference despite its comparatively large conduction-band mass. InSb serves as an established narrow-gap reference material and allows a direct comparison with previous calculations. InAs$_{x}$Sb$_{1-x}$ and Hg$_{1-x}$Cd$_x$Te are widely studied compositionally tunable infrared materials whose conduction-band effective masses can be substantially smaller than those of most conventional semiconductors. Their shallow donor levels also make low-temperature population of the conduction band more feasible.

Other $n$-type semiconductors could, in principle, be treated within the same formalism. However, many widely available alternatives possess larger conduction-band effective masses, often $M_{*}\gtrsim0.1M_0$, and substantially deeper donor levels. A larger effective mass increases the characteristic momentum of the initially occupied conduction-electron states, whereas deeper donor levels suppress their ionization at low temperature and generally require operation at higher temperatures to obtain an appreciable conduction-electron population. Both effects make the hierarchy \eqref{eq:material-oam-condition} more difficult to satisfy and, therefore, lead, already at the qualitative level, to broader photoelectron OAM distributions and poorer desired-OAM selectivity. We consequently restrict the quantitative comparison to the four materials that both offer particularly favorable band parameters and could be of direct practical relevance as photocathodes.

The theory assumes a single approximately parabolic conduction-band valley characterized by a finite, nearly constant band mass $M_{*}$. The materials are compared using the same photon TAM projection $m_\gamma$ and the same wave packet parameters $\bar k_z$, $\sigma_\perp$, and $\sigma_z$. The narrow-gap materials are studied at $T=0.5$ and $4$~K, whereas GaAs is considered at $T=2.5$ and $4$~K because of its larger donor ionization energy and the stronger freeze-out of conduction electrons at lower temperatures. The final-electron momentum components $p_{f,\perp}$ and $p_{f,z}$ are chosen near the maximum of the photoelectron detection probability.

The donor concentrations $n_d$ adopted for some of these materials were deliberately chosen near the lower end of their experimentally reported ranges in order to remain within the weak-doping regime assumed by the present model. In this regime, the donor-bound electron states are treated as spatially isolated: their wave functions have negligible overlap, and the discrete donor levels do not merge into an impurity band. The applicability of this approximation can be expressed approximately through the Mott criterion (see, for instance, \cite{Mott1961,Krieger1971}):
\begin{equation}
n_d^{1/3}\,a_d^* \lesssim 0.25,
\qquad
a_d^* = \frac{\varepsilon_0}{M_{*}/M_0}a_0,
\end{equation}
where $a_d^*$ is the effective Bohr radius of a shallow donor, {$a_0$ is the conventional Bohr radius, and $\varepsilon_0$ is a static dielectric constant of the material:
\[
    \varepsilon_0 = \varepsilon(\omega = 0).
\]}

% The corresponding transverse momentum projection of the photoelectron equals the transverse photon momentum $k_\perp^{\rm{max}}$ for the maximum of the distribution $|g(k_\perp,k_z)|^2$:
% \begin{equation}
%     p_{f,\perp} = k_\perp^{\rm max} = \sigma_\perp\sqrt{2|m_\gamma| + 1},
% \end{equation}
% while the electron longitudinal momentum projection $p_{f,z}$ is chosen from the energy conservation condition:
% \begin{equation}
%     p_{f,z} = \sqrt{2 M_0 \left(\sqrt{\bar k_z^2 + (p_{f,\perp})^2} - U_0 + T\right) - (p_{f,\perp})^2}.
% \end{equation}
For fixed detected momenta, we denote the OAM-resolved differential
probability plotted in Figs.~\ref{fig:GaAs} and
\ref{fig:semiconductor-oam-spectra} by
\begin{equation}
W_{l_e}(p_{f,z},p_{f,\perp})
\equiv
\frac{dP_{l_e}}{dp_{f,z}\,dp_{f,\perp}}.
\label{eq:oam-differential-spectrum}
\end{equation}
For each calculation, the corresponding discrete OAM distribution is
normalized as
\bea
\mathcal{P}_{l_e}=\frac{W_{l_e}}{\sum_{l'_e}W_{l'_e}},
\label{eq:oam-transition-efficiency}
\eea
This normalization isolates redistribution among OAM channels but removes information about the total photoemission probability and quantum yield. We use three complementary quantities: the mean photoelectron OAM
\bea
\langle l_e\rangle=\sum_{l_e} l_e\mathcal{P}_{l_e},
\label{eq:mean-oam}
\eea
its standard deviation,
\bea 
\Delta l_e= \left[\sum_{l_e}(l_e-\langle l_e\rangle)^2\mathcal{P}_{l_e}\right]^{1/2},
\label{eq:effective-oam-width}
\eea
and 
\bea
    \eta_m = \mathcal{P}_{l_e=m_\gamma}.
\label{eq:desired-oam-fraction}
\eea
{Here \(\eta_m\) is the fraction of detected electrons in the state with \(l_e=m_\gamma\), or \textit{the desired-OAM fraction}. Since it is defined from the normalized OAM spectrum, \(\eta_m\) characterizes OAM purity, not the absolute photoemission probability or cathode quantum efficiency.}

The calculated \(\mathcal{P}_{l_e}\) values determine the diagonal OAM populations. They do not, by themselves, determine all off-diagonal elements of the emitted-electron density matrix. Statements about coherence between different OAM channels therefore require either an explicit density-matrix calculation or an additional physical assumption about the source ensemble. 

\subsubsection{$\mathrm{GaAs}$}

Gallium arsenide is one of the most widely used semiconductor photocathode materials. GaAs-based negative electron affinity (NEA) cathodes are widely used for polarized-electron sources and can provide high quantum efficiency and low emittance \cite{Pierce1976,Sinclair2007}. Their practical operation, however, requires atomically clean surfaces, activation by Cs/O or related layers, and ultrahigh vacuum. In this work GaAs is therefore treated as a common $n$-type semiconductor.

\begin{figure*}[t]
    \centering
    \includegraphics[width=1\linewidth]{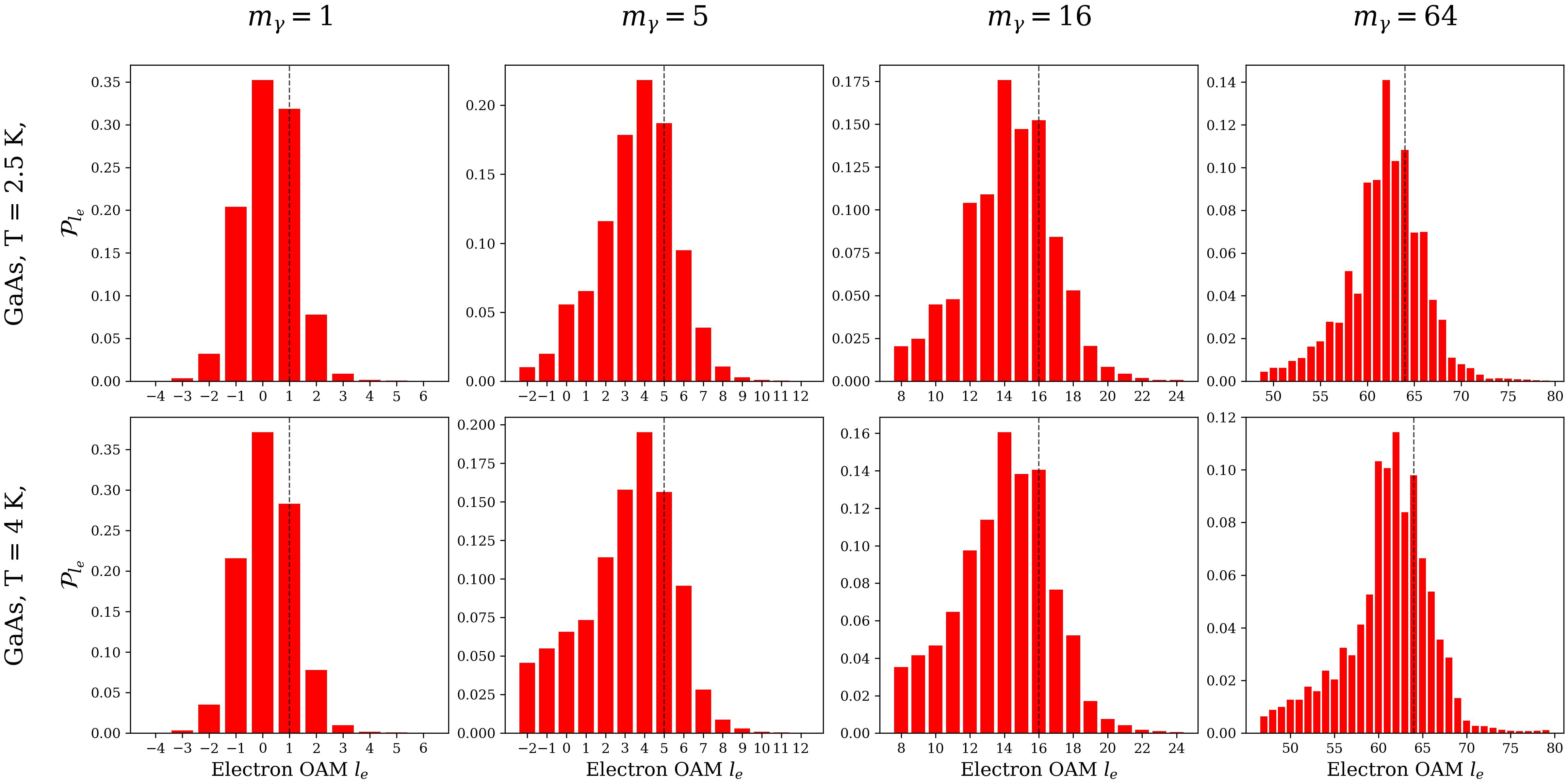}
    \caption{OAM-resolved differential photoemission probability for GaAs at
    $T=2.5$ K (top) and $T=4$ K (bottom) for an incident twisted-photon state
    with $\bar k_z=-5$ eV, $\sigma_\perp = 1$ eV and $\sigma_z = 2$ eV. The photon TAM projections are
    $m_\gamma=1,5$, $16$ and $64$. The finite thermal momentum of the initial electrons
    broadens the distribution away from the ideal selection rule $l_e=m_\gamma$. The
    effect becomes particularly strong for large $m_\gamma$.}
    \label{fig:GaAs}
\end{figure*}

Figure~\ref{fig:GaAs} shows the OAM-resolved photoelectron emission probability at $T=2.5$ and $4$ K. Under the parameters considered here, the spectra are broad already at $2.5$ K and become less selective as the temperature is raised. For bulk GaAs, the band-edge conduction-electron mass is approximately \(M_{*}\simeq0.065-0.070\,M_0\) \cite{Vurgaftman2001}, while the binding energies of shallow donors are typically about \(\varepsilon_d \simeq 5.8-7\) meV \cite{Wagner1987}. Donor concentration is chosen to be $n_d \sim 10^{16}\, \text{cm}^{-3}$, which lies in the range of practically feasible values for GaAs $n_d \sim 10^{15} - 10^{17}\, \text{cm}^{-3}$ \cite{Kobori1995, Zakharin2010}. Donor freeze-out therefore suppresses the conduction-electron population at sufficiently low temperature. Heating ionizes a larger fraction of the donors, but simultaneously increases the characteristic transverse momentum of the initial electrons and broadens the emitted-electron OAM spectrum. Within the present \(n\)-type conduction-electron model, GaAs is unfavorable for selective OAM transfer because its relatively large effective mass broadens the initial momentum distribution, while its comparatively large donor binding energy strengthens carrier freeze-out at low temperature. Increasing the donor concentration can improve the available electron population, but it also raises the Fermi momentum and further reduces the OAM selectivity. This conclusion applies to the model considered here and does not directly describe the usual activated NEA operating regime of practical GaAs photocathodes.

\subsubsection{$\mathrm{InSb}$}

Indium antimonide provides the reference case for the present comparison. It is a technologically established narrow-gap semiconductor used in infrared detectors and magnetotransport devices \cite{Vurgaftman2001,Shiomi2016}. Among conventional narrow-gap semiconductors, InSb has an exceptionally small band-edge conduction-electron mass, $M_{*} \approx 0.014\,M_0$ \cite{Vurgaftman2001}, and very shallow donor levels, $\varepsilon_d \approx 7\times 10^{-4}$ eV \cite{IoffeInSb}, which reduce the characteristic momentum of the thermally occupied conduction-electron states. This helps to satisfy the momentum hierarchy \(p_{i,\perp} \ll p_{f,\perp} \simeq k_\perp\) underlying the approximate selection rule \(l_e = m_\gamma\), and, therefore, makes InSb a natural benchmark for achieving high-fidelity transfer of angular momentum from the incident twisted photon to the emitted photoelectron. Shallow donor concentration is $n_d \sim 10^{14}\, \text{cm}^{-3}$ \cite{Tokumoto1982}.

The first pair of rows in Fig.~\ref{fig:semiconductor-oam-spectra} shows the calculated InSb spectra at $T=0.5$ and $4$ K. At $0.5$ K, the $m_\gamma=1$ distribution is strongly concentrated in the desired OAM channel: $\eta_1=87\%$, $\langle l_e\rangle=0.9$, and $\Delta l_e=0.4$. For $m_\gamma=5$ and 16, the mean electron OAM remains close to the incident-photon TAM projection, with $\langle l_e\rangle=4.8$ and 15.7, while the desired-OAM fractions are 55\% and 35\%, respectively. The nonmonotonic change of $\eta_m$ between these two values shows that the detailed spectrum is governed not only by a single thermal broadening scale but also by the structure of the transition amplitude. For $m_\gamma=64$, the mean remains large, $\langle l_e\rangle=63.0$, but the desired OAM channel contains only 16\% of the normalized probability and the rms width reaches $\Delta l_e=2.5$. Thus, a nearly correct mean OAM does not imply preparation of a nearly pure $l_e=64$ state.

At $4$ K, all four InSb distributions broaden and become substantially less selective. The mean values decrease to 0.5, 4.2, 14.7, and 62.5 for $m_\gamma=1$, 5, 16, and 64, respectively, whereas the rms widths increase to 0.9, 1.4, 2.1, and 3.1. The corresponding desired-OAM fractions fall to 45\%, 24\%, 18\%, and 8\%. The downward displacement of the means indicates preferential redistribution toward channels with $l_e<m_\gamma$. The $m_\gamma=64$ case is especially illustrative: its mean OAM is still close to the photon TAM projection, but only 8\% of the detected electrons occupy the desired OAM channel.

\begin{table*}[t]
\caption{
Comparison of OAM-transfer metrics for InSb,
InAs$_{0.34}$Sb$_{0.66}$,
and Hg$_{0.76}$Cd$_{0.24}$Te at $T=0.5$ and $4$ K.
{The mean OAM $\langle l_e\rangle$, rms width $\Delta l_e$, and desired-OAM fraction $\eta_m$ are defined in \eqref{eq:mean-oam}, \eqref{eq:effective-oam-width}, and \eqref{eq:desired-oam-fraction}, respectively.}
}
\label{tab:oam-metrics-all}

\centering
\footnotesize
\setlength{\tabcolsep}{2.5pt}
\renewcommand{\arraystretch}{1.3}

\begin{ruledtabular}
\begin{tabular}{c|cccc|cccc|cccc}

& \multicolumn{4}{c|}{InSb}
& \multicolumn{4}{c|}{InAs$_{0.34}$Sb$_{0.66}$}
& \multicolumn{4}{c}{Hg$_{0.76}$Cd$_{0.24}$Te}
\\

$m_\gamma$
& $T$ (K) & $\langle l_e\rangle$ & $\Delta l_e$ & $\eta_m$ (\%)
& $T$ (K) & $\langle l_e\rangle$ & $\Delta l_e$ & $\eta_m$ (\%)
& $T$ (K) & $\langle l_e\rangle$ & $\Delta l_e$ & $\eta_m$ (\%)
\\
\hline

1
& 0.5 & 0.9 & 0.4 & 87
& 0.5 & 1.0 & 0.4 & 91
& 0.5 & 1.0 & 0.3 & 91
\\

&
  4.0 & 0.5 & 0.9 & 45
& 4.0 & 0.6 & 0.9 & 55
& 4.0 & 0.6 & 0.8 & 56
\\
\hline

5
& 0.5 & 4.8 & 0.8 & 55
& 0.5 & 4.9 & 0.7 & 66
& 0.5 & 4.9 & 0.7 & 66
\\

&
  4.0 & 4.2 & 1.4 & 24
& 4.0 & 4.4 & 1.2 & 28
& 4.0 & 4.4 & 1.2 & 28
\\
\hline

16
& 0.5 & 15.7 & 1.2 & 35
& 0.5 & 15.7 & 1.2 & 36
& 0.5 & 15.7 & 1.2 & 37
\\

&
  4.0 & 14.7 & 2.1 & 18
& 4.0 & 15.0 & 1.9 & 16
& 4.0 & 15.0 & 1.9 & 17
\\
\hline

64
& 0.5 & 63.0 & 2.5 & 16
& 0.5 & 63.4 & 2.1 & 17
& 0.5 & 63.3 & 2.2 & 17
\\

&
  4.0 & 62.5 & 3.1 & 8
& 4.0 & 62.7 & 3.0 & 15
& 4.0 & 62.8 & 3.0 & 14
\\

\end{tabular}
\end{ruledtabular}
\end{table*}

\subsubsection{$\mathrm{InAs}_{x}\mathrm{Sb}_{1-x}$}

InAs$_{x}$Sb$_{1-x}$ is a compositionally tunable narrow-gap III--V alloy widely studied for infrared detection \cite{Rogalski2020}. Its conduction-band mass can remain small, while its band gap, dielectric response, and carrier properties can be adjusted through the As/Sb ratio. The present calculations use InAs\(_{0.34}\)Sb\(_{0.66}\), a composition close to the minimum-gap region of the alloy system \cite{Rogalski2020}. We adopt \(M_{*}=0.009\,M_0\), consistent with the band-edge masses reported for Sb-rich and near-minimum-gap InAs$_{x}$Sb$_{1-x}$ alloys \cite{Vurgaftman2001}. The donor concentration was set to $n_d=10^{14}\,\mathrm{cm}^{-3}$ as a weak-doping model parameter. Background dopant or carrier concentrations in the $10^{14}$--$10^{15}\,\mathrm{cm}^{-3}$ range have been experimentally reported for high-quality InAsSb epilayers \cite{Morath2023}. The shallow-donor ionization energy \(\varepsilon_d=6\times10^{-4}\) eV is estimated within the hydrogenic effective-mass model. Finally, the crystal--vacuum offset \(U_0=4.7\) eV is based on the composition-dependent electron-affinity interpolation tabulated for InAs$_{x}$Sb$_{1-x}$ \cite{IoffeInAsSb}. Thus, these quantities should be regarded as a consistent parameter set for the calculation rather than as universal properties of every InAs\(_{0.34}\)Sb\(_{0.66}\) sample.

The middle pair of rows in Fig. \ref{fig:semiconductor-oam-spectra} demonstrates the strong localization of the spectra around the desired channels for $m_\gamma=1$ and 5 at $T=0.5$ K. Their desired-OAM fractions are 91\% and 66\%, with mean OAM values of 1.0 and 4.9 and rms widths of 0.4 and 0.7, respectively. For $m_\gamma=16$, the mean remains close to the incident value, $\langle l_e\rangle=15.7$, but $\eta_{16}$ decreases to 36\% and $\Delta l_e=1.2$. For $m_\gamma=64$, the spectrum has $\langle l_e\rangle=63.4$, $\Delta l_e=2.1$, and $\eta_{64}=17\%$.

At $4$ K, the mean values decrease to 0.6, 4.4, 15.0, and 62.7 for $m_\gamma=1$, 5, 16, and 64, while the rms widths increase to 0.9, 1.2, 1.9, and 3.0. The desired-OAM fractions are 55\%, 28\%, 16\%, and 15\%, respectively. The modest decrease of $\eta_{64}$ from 17\% to 15\% should not be interpreted as thermal stability of a pure OAM mode: over the same temperature interval the rms width grows from 2.1 to 3.0 and the mean moves farther below 64. This behavior confirms that $\eta_m$ must be considered together with $\langle l_e\rangle$ and $\Delta l_e$.

Under the present parameter choices, InAs\(_{0.34}\)Sb\(_{0.66}\) retains better desired-OAM selectivity than InSb at $4$ K for $m_\gamma=1$, 5, and 64, although the difference is not universal for every $m_\gamma$. It therefore represents a useful narrow-gap material for studying OAM transfer in the surface photoelectric effect.

\subsubsection{$\mathrm{Hg}_{1-x}\mathrm{Cd}_{x}\mathrm{Te}$}

Hg\(_{1-x}\)Cd\(_x\)Te is a compositionally tunable narrow-gap semiconductor extensively employed in infrared detectors \cite{Rogalski2005}. Near \(x=0.24\), its small band gap is accompanied by a low conduction-band effective mass. The values \(M_{*}=0.009\,M_0\) and \(\varepsilon_0=17.1\) follow from the composition- and temperature-dependent parameterizations summarized in \cite{Rogalski2005}. The corresponding hydrogenic estimate gives the shallow-donor ionization energy \(\varepsilon_d\simeq4\times10^{-4}\) eV. {The donor concentration \(n_d\sim10^{13}\,\mathrm{cm}^{-3}\) is taken as a representative weak-doping value, while \(U_0=4.2\) eV is adopted as the crystal–vacuum energy offset for the assumed surface.} Because the Hg\(_{0.76}\)Cd\(_{0.24}\)Te band structure is strongly nonparabolic, the validity of the constant-mass approximation must be checked over the occupied energy range included in the calculation.

\begin{figure*}[t]
    \centering
    \includegraphics[width=1\linewidth]{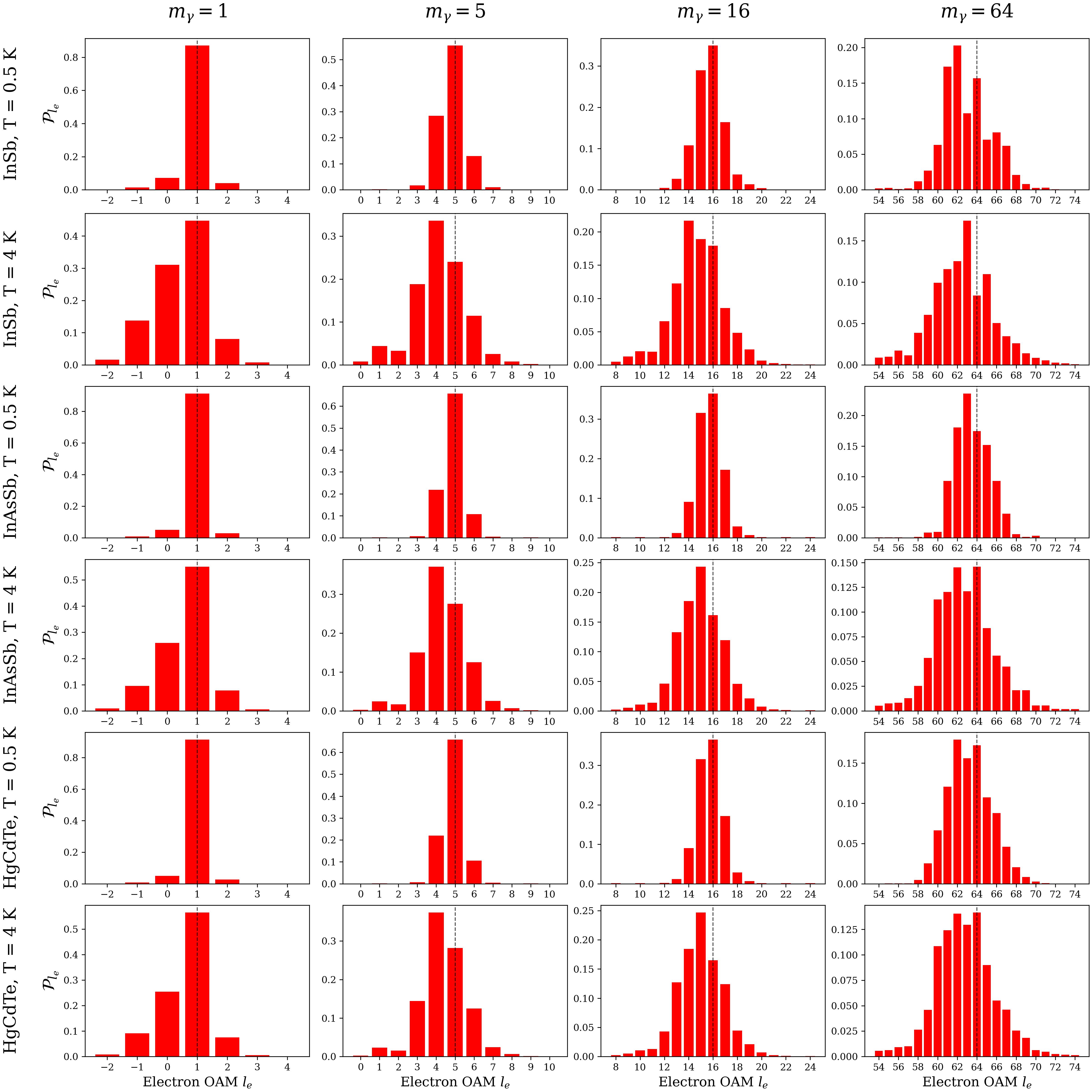}
    \caption{OAM-resolved normalized differential photoemission probability for InSb, InAs$_{0.34}$Sb$_{0.66}$, and Hg$_{0.76}$Cd$_{0.24}$Te. The four columns correspond to the total angular-momentum projections of the incident twisted photon, $m_\gamma=1$, $5$, $16$, and $64$, respectively, for $\bar k_z = -5$ eV, $\sigma_\perp = 1$ eV, $\sigma_z = 2$ eV. For each material, the upper and lower rows show the photoelectron OAM distributions at $T=0.5$ K and $T=4$ K, respectively. The pairs of rows from top to bottom correspond to InSb, InAs$_{0.34}$Sb$_{0.66}$, and Hg$_{0.76}$Cd$_{0.24}$Te. The dashed vertical lines indicate the desired OAM-transfer channel $l_e=m_\gamma$.}
    \label{fig:semiconductor-oam-spectra}
\end{figure*}

The lower pair of rows in Fig. \ref{fig:semiconductor-oam-spectra} shows that at $T=0.5$ K, Hg\(_{0.76}\)Cd\(_{0.24}\)Te gives the narrowest $m_\gamma=1$ distribution among the cases considered: $\eta_1=91\%$, $\langle l_e\rangle=1.0$, and $\Delta l_e=0.3$. For $m_\gamma=5$, the mean is 4.9, the desired-OAM fraction is 66\%, and the rms width is 0.7. At $m_\gamma=16$, $\langle l_e\rangle=15.7$, $\eta_{16}=37\%$, and $\Delta l_e=1.2$. For $m_\gamma=64$, the mean remains close to the incident value, $\langle l_e\rangle=63.3$, but only 17\% of the normalized probability lies in the $l_e=64$ channel and the rms width is 2.2.

Raising the temperature to $4$ K reduces the selectivity. The mean values become 0.6, 4.4, 15.0, and 62.8, and the rms widths increase to 0.8, 1.2, 1.9, and 3.0 for $m_\gamma=1$, 5, 16, and 64, respectively. The corresponding desired-OAM fractions are 56\%, 28\%, 17\%, and 14\%. These results are very close to those obtained for InAs\(_{0.34}\)Sb\(_{0.66}\) at the same temperature. For $m_\gamma=64$, the relatively small change of $\eta_{64}$, from 17\% to 14\%, again does not imply thermal stability of a pure mode, because the rms width increases from 2.2 to 3.0 and the mean shifts farther below the desired value.

\section{discussion}

\subsection{General trends}

The results reveal the common behavior shared by the narrow-gap semiconductors considered in this work. Increasing the photocathode temperature from $0.5$ to $4$~K broadens the photoelectron OAM distributions, increases their rms widths $\Delta l_e$, and generally reduces the desired-OAM fraction $\eta_m$. This loss of selectivity becomes more pronounced as the TAM projection, $m_\gamma$, of the incident photon is increased. Moreover, the mean electron OAM shifts below the desired value $m_\gamma$ at $4$~K for all the considered materials. {Within the current model, twisted-photon photoemission therefore cannot generally be regarded as the direct preparation of a pure electron OAM eigenstate with $l_e=m_\gamma$.}

The mean OAM alone is not a reliable measure of the modal purity of the emitted electrons. For example, at $T=0.5$~K and $m_\gamma=64$, the mean values for the three narrow-gap materials remain between $63.0$ and $63.4$, {although only $16-17\%$ of the electrons are found in the desired $l_e=64$ state. At $4$~K, the corresponding mean values remain between $62.5$ and $62.8$, whereas the fraction in that desired state decreases to $8-15\%$.} Thus, a mean electron OAM close to the photon TAM projection can coexist with a broad distribution over neighboring OAM channels. The quantities $\langle l_e\rangle$, $\Delta l_e$, and $\eta_m$ must therefore be considered together when evaluating the quality of OAM transfer.

The broadening originates from the finite momentum distribution of the initially occupied electronic states in the crystal. The photoemission probability is averaged over this thermal ensemble \eqref{eq:fermi-dirac-density-matrix}, and electrons with different initial transverse momenta acquire different corrections to the ideal selection rule $l_e=m_\gamma$, as seen from \eqref{eq:oam-amplitude-expansion}. {This, therefore, gives an incoherent mixture of different OAM components in the final state, rather than a coherent superposition of OAM modes. The quantity $\Delta l_e$ thus characterizes the width of this OAM distribution. A coherent superposition would imply nonzero off-diagonal elements of the emitted-electron density matrix, which are not described by the present calculation.} Efficient OAM transfer requires the characteristic transverse momentum $p_{i,\perp}$ of the initial electrons to remain small compared with the transverse momentum width of the incident photon packet,

\begin{equation}
p_{i,\perp} \ll \sigma_\perp.
\end{equation}

This condition is favored by low photocathode temperatures, small conduction-band effective masses, and shallow donor levels. It also shows that the transfer efficiency is determined jointly by the material parameters, temperature, and momentum-space structure of the incident twisted-photon wave packet.

Under the parameter choices adopted here, InAs$_{0.34}$Sb$_{0.66}$ and Hg$_{0.76}$Cd$_{0.24}$Te exhibit closely comparable OAM distributions and are generally more selective than InSb, primarily because of their smaller conduction-band effective masses and shallow donor ionization energies. GaAs is less favorable within the present bulk $n$-type conduction-electron model because its larger effective mass produces a wider thermal momentum distribution. These results should not, however, be interpreted as a universal ranking of photocathode materials. A realistic comparison must also account for the absolute photoelectron yield, donor compensation and ionization, surface preparation, crystal--vacuum interface properties, and the validity range of the parabolic constant-mass approximation.

{A finite OAM width is not necessarily a drawback. Coherent occupation of several OAM modes can provide a high-dimensional degree of freedom, as demonstrated for OAM-entangled photons \cite{revMair2001}, and wave-packet phases and transverse coherence can play an essential role in scattering observables \cite{revKarlovets2017JHEP,revKarlovetsSerbo2020}. Thus, if photoemission preserves off-diagonal coherence between OAM components, or if the emitted electron becomes entangled with another degree of freedom or another particle, a controlled nonzero $\Delta l_e$ may become a useful resource rather than merely a loss of mode purity. However, the present OAM populations $\mathcal P_{l_e}$ do not establish such coherence or entanglement. Testing this possibility requires a density-matrix calculation and correlation-sensitive measurements.}

The incomplete OAM selectivity of semiconductor photoemission suggests two complementary directions for further development. The first is to consider materials with more favorable low-energy band structures, including Dirac and Weyl semimetals, although their description requires an extension beyond the parabolic effective-mass approximation. The second is to separate the emitted OAM components after photoemission and postselect the desired channel using an electron OAM sorter. These possibilities are discussed below.

\subsection{Dirac and Weyl semimetals}

Dirac and Weyl semimetals constitute a potentially favorable, although
experimentally less feasible, class of photocathode materials. High-quality
crystals and thin films with controlled stoichiometry, carrier density, chemical
potential, and surface termination are more difficult to fabricate than established
group III--V and II--VI semiconductors, and these materials are not yet standard
photocathode platforms. Nevertheless, their low-energy band structure can be
advantageous for transferring angular momentum from a twisted photon to an
electron.

Near a Weyl node with energy $E_W$, the effective Hamiltonian can be written schematically as
\begin{equation}
    H_{\chi}(\bm q)=E_W+\chi\sum_{a=x,y,z} v_a\sigma_a q_a,
\end{equation}
where $\bm q$ is the crystal momentum measured from the node, $v_a$ are the
generally anisotropic Fermi velocities, $\sigma_a$ are Pauli matrices, and
$\chi=\pm1$ is the node
chirality. A Dirac node contains two degenerate Weyl sectors of opposite
chirality. The corresponding dispersion is
\begin{equation}
    E_{\pm}(\bm q)=E_W\pm
    \left(\sum_{a=x,y,z} v_a^2q_a^2\right)^{1/2}.
\end{equation}
Such a spectrum has no finite, constant band mass that would reproduce it in
the parabolic form $p^2/(2M_{*})$. For an isotropic cone, however, the cyclotron
or dynamical mass at energy $E$ can be introduced as
\begin{equation}
    m_c(E)=\frac{|E-E_W|}{v_F^2},
\end{equation}
where $v_F$ is the isotropic Fermi velocity. This mass tends to zero as the chemical potential approaches the node or, equivalently,
the Fermi momentum $p_F=|\mu-E_W|/v_F$ becomes small. This reduction in the characteristic momentum of the initial electrons makes it easier to satisfy the condition $p_{i,\perp}\ll p_{f,\perp}\sim k_\perp$ and thereby improve OAM transfer.

Linear dispersion alone does not guarantee a useful photocathode. Native
doping or defects can displace the chemical potential far from the node and
increase $p_F$, whereas the density of bulk states vanishes near an ideal
three-dimensional node, potentially reducing the available photoelectron
current. Multiple nodes, anisotropy and tilt of the cones, spinor or pseudospin
selection rules, surface states, band bending, and the crystal--vacuum boundary
can also modify the photoemission amplitude.

The present formalism assumes a scalar electron state with a parabolic bulk
dispersion, $E=p^2/(2M_{*})$, and velocity $\bm v=\bm p/M_{*}$. It cannot be
extended to a Dirac or Weyl semimetal by simply setting $M_{*}=0$. A consistent
treatment must replace the bulk Hamiltonian and velocity operator, retain the
multicomponent Bloch spinors, recalculate the equilibrium carrier distribution,
and derive the interface matching conditions and photoemission matrix element
for the relevant nodes. Cd$_3$As$_2$ and Na$_3$Bi, as representative three-dimensional Dirac semimetals, together with critical-composition Hg$_{1-x}$Cd$_x$Te (\(x\simeq0.15\text{--}0.17\)), which realizes a zero-gap Kane system \cite{Teppe2016}, constitute well-studied model materials for future theoretical and experimental investigations of twisted-photon photoemission and OAM transfer. Their suitability as practical photocathodes, however, remains to be established.

\subsection{Preparation of OAM-selected electron states with mode sorters}
\label{subsec:oam-sorter}

The photocathode does not need to emit electrons exclusively in the desired-OAM state. If several OAM components are produced, a downstream OAM sorter can map them onto different spatial channels. An aperture can then transmit the desired component \(l_e=m_\gamma\) and block the others. This makes it possible to obtain an electron beam with high OAM selectivity, although at the cost of reducing the total current. In the ideal case, the fraction of electrons available for postselection is limited by the desired-OAM fraction \(\eta_m\) calculated above. Such postselection becomes particularly useful at finite temperatures and for photons with large \(m_\gamma\), when the emitted-electron OAM distribution broadens.

Sorting alone does not increase the purity of the electron state: it only separates its OAM components spatially. Purification is achieved conditionally by transmitting one output channel. Moreover, selection of a definite \(l_e\) does not necessarily select a unique radial mode, electron energy, or transverse momentum. The resulting beam should therefore be described as OAM-selected rather than as a completely pure transverse quantum state.

{A magnetic-quadrupole Laguerre-Gaussian--Hermite-Gaussian (LG--HG) converter \cite{Schattschneider2012,Kramberger2019,Schachinger2021} provides one possible method for analysing electron vortex modes.} It uses astigmatic focusing to transform a Laguerre--Gaussian mode into a Hermite--Gaussian mode of the same total transverse order,
\begin{equation}
N=2n+|l_e| = s_x + s_y,
\label{eq:lg-hg-order}
\end{equation}
where \(n\) is the radial index of the Laguerre--Gaussian mode and $s_x$ and $s_y$ are the transverse mode indices of the Hermite--Gaussian mode \cite{Schattschneider2012,Kramberger2019,Schachinger2021}. This method is not well suited to the present problem because the photoemitted electrons occupy a mixture of OAM, radial, and transverse-momentum states. The corresponding HG-like intensity patterns would overlap, and different \((p,l)\) states can have the same value of \(N\). The converter therefore does not map each value of \(l_e\) onto a unique spatial channel.

\subsubsection{Electrostatic log-polar sorter}

A log-polar conformal sorter provides a more direct method for separating the OAM components. It implements the coordinate transformation
\begin{equation}
u=a\ln(\rho/b),
\qquad
v=a\varphi,
\label{eq:logpolar-map}
\end{equation}
under which the azimuthal phase of an OAM eigenstate becomes a linear phase ramp,
\begin{equation}
e^{il_e\varphi}
\longrightarrow
e^{i(l_e/a)v}.
\label{eq:oam-linear-ramp}
\end{equation}
Here \((\rho,\varphi)\) are the input polar coordinates, \((u,v)\) are the unwrapped coordinates, \(a\) sets the transverse scale of the transformation, and \(b\) is a reference radial scale. A subsequent Fourier transformation maps different phase gradients, and hence different values of \(l_e\), onto different transverse positions. Log-polar sorting of electron OAM has been demonstrated using nanofabricated holograms and electrostatic phase elements \cite{Grillo2017,Tavabi2021}. In the electrostatic implementation, an OAM resolution of approximately \(1.5\hbar\) was achieved for 300-keV electrons \cite{Tavabi2021}.

The finite extent of the unwrapped electron wave gives each output channel a nonzero width, so neighboring OAM peaks can overlap. The resolution can be improved by a fan-out transformation, which produces \(N\) phase-corrected copies of each unwrapped component. These copies form a wider effective aperture and reduce the Fourier-space width, $\Delta k^{(N)}$, approximately as
\begin{equation}
\Delta k^{(N)}
\sim
\frac{1}{N}\Delta k^{(1)},
\label{eq:fanout-oam-resolution}
\end{equation}
while leaving the separation between adjacent OAM peaks approximately unchanged \cite{Mirhosseini2013,Ruffato2018,Rosi2021,RosiQSORT}.

Neither the log-polar transformation nor the separation of the output intensities requires mutual coherence between different OAM components. For an incoherent OAM mixture, each component is transformed independently, and the corresponding detector intensities add. The fan-out stage requires only sufficient transverse coherence within each individual component for its copies to interfere with controlled relative phases.

For state preparation, an aperture can be placed at the output position corresponding to \(l_e=m_\gamma\). If required, an inverse log-polar transformation can then restore a vortex-like transverse profile. The principal limitations are cross-talk between neighboring channels, losses during postselection, sensitivity to alignment and aberrations, and the need for specially fabricated phase elements. Furthermore, existing electron implementations have primarily operated at transmission-electron-microscope energies, so their adaptation to the few-eV photoelectrons considered here requires a dedicated electron-optical design.

\subsubsection{Alternative OAM analyzers}

{A Dammann vortex analyzer \cite{Noguchi2019} performs several predetermined OAM projections simultaneously.} A multiplexed grating sends the incident electron wave into multiple diffraction orders, each containing the conjugate phase of a selected OAM state. If the incident beam contains the corresponding component, its azimuthal phase is cancelled and a compact correlation peak appears in that diffraction order. Electron OAM components from \(-10\hbar\) to \(+10\hbar\) have been measured using this method \cite{Noguchi2019}. The analyzer is compact and suitable for parallel measurements, but it probes only the finite set of OAM values encoded in the grating, divides the available intensity among several diffraction orders, and has a response that depends on the radial profile of the incident beam.

A Stern--Gerlach-like OAM analyzer instead uses the interaction of the electron orbital magnetic moment with a cylindrically symmetric inhomogeneous magnetic field \cite{Harvey2017}. Different OAM components acquire different focal lengths and can, therefore, be distinguished through their radial distributions. This noninterferometric method can operate on incoherent mixtures and, because it preserves the azimuthal phase, may not require an inverse transformation after postselection. Its main limitations are the weak OAM-dependent magnetic dispersion expected for realistic fields and its sensitivity to electron energy, beam size, divergence, and radial-mode content. The lower kinetic energies considered in the present work may enhance the OAM-dependent focusing, but a separate low-energy electron-optical analysis is required to establish whether this method can provide sufficient resolution.

\subsection{Implications for a proof-of-principle experiment}

{The RF-photoinjector test bench under development at JINR presently uses a copper photocathode illuminated by a $260$--$266$~nm UV laser \cite{revDyatlov2025BeamDynamics,Dyatlov2026}. Copper is convenient technologically, but it is unfavorable for selective OAM transfer in the sense of \eqref{eq:material-oam-condition}. In a free-electron estimate, Cu has a conduction-electron density $n_e\simeq8.5\times10^{22}\,\mathrm{cm}^{-3}$ and a Fermi energy $E_F\simeq7$~eV \cite{revAshcroftMermin1976}, corresponding to a Fermi momentum $p_F\simeq2.7$~keV. For an isotropic filled Fermi sphere, $\sqrt{\langle p_{i,\perp}^2\rangle}=\sqrt{2/5}\,p_F\simeq1.7$~keV, which is 3 orders of magnitude larger than the eV-scale transverse momentum widths of a DUV-photon. Cooling does not remove this limitation because copper remains strongly degenerate even for the temperatures of $\sim 1$ K. Thus, a copper cathode is expected to produce a much broader OAM distribution than the weakly doped narrow-gap semiconductors treated above.}

For a first experimental demonstration, the most realistic strategy is to use a material that provides a sufficiently large desired-OAM fraction at small or moderate photon TAM projection, rather than attempting very large $m_\gamma$ immediately. In the calculations presented above, the cases $m_\gamma=1$ and $5$ generally produce narrower OAM spectra, require a smaller sorter field of view, and retain a larger fraction of the emitted electrons in the desired $l_e=m_\gamma$ channel. Such conditions therefore appear more suitable for a proof-of-principle experiment.

The experiment should measure both the total photoelectron yield and the normalized OAM distribution $\mathcal{P}_{l_e}$. A material can produce a comparatively narrow OAM distribution while yielding an impractically small electron current, whereas a high-yield photocathode may generate a broader mixture of OAM states. A log-polar sorter supplemented by fan-out enhancement could directly test the calculated $\mathcal{P}_{l_e}$ and subsequently postselect the $l_e=m_\gamma$ component, while a Dammann-vortex analyzer programmed for a narrow interval around $m_\gamma$ could provide an independent measurement.

An additional experimental constraint originates from the longitudinal and transverse momentum widths of the incident photon wave packet. In the present calculations, $\sigma_z\equiv\sigma_{\parallel}$ and $\sigma_\perp$ are typically of order $1$~eV, with values up to several eV considered in some parameter sets. Experimentally available ultraviolet sources span a broad range of longitudinal widths. {Twisted photons have been generated at $266$~nm with a spectral width of $1.02$~nm \cite{Chaitanya2016},} corresponding to an energy width of approximately $1.8\cdot10^{-2}$~eV. In ultrafast photoemission experiments, UV bandwidths between $2$ and $10$~nm at the same wavelength have also been produced \cite{Aidelsburger2010}, corresponding approximately to $3.5\cdot10^{-2}$ -- $1.8\cdot10^{-1}$~eV. Thus, characteristic widths in the $10^{-2}$--$10^{-1}$~eV range are directly supported by experiment.

Much broader DUV spectra are also possible. The work \cite{Galli2019} demonstrated $1.9$-fs DUV pulses with a spectrum extending approximately from $210$ to $340$~nm, corresponding to a photon energy span exceeding $2$~eV. This demonstrates that eV-scale spectral width is experimentally accessible, although the full spectral span should not be identified directly with the rms Gaussian width $\sigma_z$. For the longitudinal Gaussian distribution \eqref{eq:LG-momentum-4.6}, the root-mean-squared temporal width of the packet is
\begin{equation}
\sigma_t=\frac{\hbar}{2\sigma_z},
\end{equation}
which correspond to the following intensity full-width-at-half-maximum (FWHM) pulse duration
\begin{equation}
\Delta t_{\mathrm{FWHM}}
=
\sqrt{2\ln2}\,
\frac{\hbar}{\sigma_z}.
\end{equation}
Consequently, $\sigma_z=1$~eV corresponds to an intensity FWHM duration of $\approx 0.78$~fs, while $\sigma_z=3$~eV - to $\approx 0.26$~fs. The largest longitudinal widths considered here should therefore be regarded as demanding, although modern sub-2-fs DUV experiments already approach the relevant energy scale.

The transverse requirement is more favorable. With the convention used in \eqref{eq:LG-momentum-4.6}, the corresponding transverse waist parameter \(w_0\) is
\begin{equation}
w_0=\frac{\hbar c}{\sigma_\perp},
\end{equation}
so that $\sigma_\perp=1$~eV corresponds to $w_0\simeq200$~nm. Diffraction-limited focusing of $266$-nm radiation to a transverse scale of approximately $150$~nm has recently been demonstrated using a high-numerical-aperture DUV metalens \cite{Abdelraouf2026}. It shows that transverse localization on the scale implied by $\sigma_\perp\sim1$~eV is experimentally accessible. For an $n=0$ Laguerre--Gaussian mode, the intensity maximum occurs at
\begin{equation}
\rho_{\max}
=
w_0\sqrt{\frac{|l_\gamma|}{2}},
\end{equation}
so increasing optical OAM, $l_\gamma$, also increases the transverse extent of the illuminated region. The principal experimental challenge is therefore the simultaneous realization of a tightly focused, sufficiently coherent, broadband UV vortex pulse with a well-controlled OAM state.

The complete photoemission-and-sorting scheme provides a conceptually different route to vortex-electron generation from established electron-optical methods. Electron vortex beams have already been produced using spiral phase plates \cite{Uchida2010}, nanofabricated diffraction holograms \cite{Verbeeck2010,McMorran2011}, magnetic-needle ``chopstick'' phase plates \cite{Beche2014,Beche2017}, and tunable electrostatic spiral phase plates \cite{Tavabi2022}. These techniques are experimentally mature and can provide high efficiency, high mode purity, and very large OAM.

Its main potential advantage is that the OAM is introduced already at the electron-generation stage. The electron inherits angular momentum, as well as energy and linear momentum components and their widths, from the structured optical field during photoemission rather than acquiring it subsequently from an electron-optical phase element. The photon TAM projection and other photon characteristics therefore become external control parameters of the source, which can in principle be changed optically without replacing an electron hologram or phase plate. Moreover, because the electrons are generated by a pulsed optical field, their emission is intrinsically synchronized with the driving laser, making the approach naturally compatible with ultrafast pump--probe configurations.

The photocathode does not need to produce a perfectly pure $l_e=m_\gamma$ state. Instead, it generates an OAM distribution $\mathcal{P}_{l_e}$, while the sorter spatially resolves its components and postselects the desired one. This makes it possible to control the trade-off between transmitted current and OAM purity. At the same time, measuring $\mathcal{P}_{l_e}$ as a function of temperature, carrier density, photon TAM projection, \(m_\gamma\), and photocathode material directly probes the microscopic transfer of angular momentum in photoemission, whereas conventional holograms and phase plates simply impose a prescribed phase on an already formed electron beam.

{The JINR program provides a concrete route for testing these ideas: structured DUV light with large OAM has already been integrated into the RF-photoinjector laser line \cite{Dyatlov2026}, while the test-bench beam-dynamics studies explicitly target relativistic vortex-electron generation \cite{revDyatlov2025BeamDynamics}. If OAM-carrying photoelectrons are subsequently accelerated, the preservation and manipulation of their angular momentum in accelerator fields become relevant, as discussed in \cite{revKarlovets2026PRL}. Besides the near-eigenstate regime, experiments could deliberately explore electron states with a finite OAM width and test whether intermode coherence survives emission and acceleration.}

The price for this additional functionality is substantial experimental complexity: a suitable photocathode, a stable structured UV source, tight focusing and transverse coherence, operation in a material regime favorable for OAM transfer, and a downstream OAM analyzer are all required, while postselection inevitably reduces the usable current. Therefore, the proposed scheme should be viewed as a complementary vortex-electron source architecture in which electron OAM is optically controlled at emission and subsequently purified by mode sorting.

\section{Conclusion}

We have studied the transfer of the angular-momentum projection of a twisted photon to photoelectrons emitted from weakly doped \(n\)-type GaAs, InSb, \(\mathrm{InAs}_{0.34}\mathrm{Sb}_{0.66}\), and \(\mathrm{Hg}_{0.76}\mathrm{Cd}_{0.24}\mathrm{Te}\). Within the single-valley parabolic-band model, selective population of the channel \(l_e=m_\gamma\) requires the characteristic transverse momentum of the initial electrons to be small compared with the transverse momentum scales of the photon wavepacket. This condition is favored by low temperature, a small conduction-band effective mass, and shallow donor levels.

Under the adopted parameters, \(\mathrm{InAs}_{0.34}\mathrm{Sb}_{0.66}\) and \(\mathrm{Hg}_{0.76}\mathrm{Cd}_{0.24}\mathrm{Te}\) produce OAM distributions comparable to or narrower than those obtained for InSb, whereas GaAs is less selective because of its larger effective mass and stronger donor freeze-out. The comparison should not be regarded as a universal ranking of photocathodes, since the present calculations do not determine the absolute photoemission yield and do not describe activated negative-electron-affinity cathodes.

The mean electron OAM alone does not characterize the selectivity of the transfer. At \(T=0.5\)~K and \(m_\gamma=64\), the mean OAM remains between \(63.0\) and \(63.4\) for the three narrow-gap materials, although only \(16\)--\(17\%\) of the normalized probability occupies the desired OAM channel. At \(4\)~K, this fraction decreases to \(8\)--\(15\%\), while the rms width reaches approximately \(3\). The mean OAM, rms width, and desired-OAM fraction must therefore be considered together.

A proof-of-principle experiment is most realistic for small or moderate photon TAM projections, such as \(m_\gamma=1\) or \(5\), and would require a cryogenic photocathode together with a broadband, tightly focused DUV vortex pulse. If several OAM components are emitted, a downstream log-polar sorter and a mode-selective aperture can isolate the desired \(l_e=m_\gamma\) channel, at the cost of reducing the electron current. Dirac and Weyl semimetals may provide more favorable carrier-momentum scales, but their quantitative treatment requires a multiband photoemission theory and cannot be obtained by simply setting \(M_*=0\) in the present model.

{The JINR RF-photoinjector project \cite{revDyatlov2025BeamDynamics,Dyatlov2026} offers a direct experimental setting for testing twisted-photon photoemission, OAM analysis, and subsequent acceleration, for which the OAM dynamics discussed in~\cite{revKarlovets2026PRL} becomes relevant. The presently used copper cathode is unlikely to provide narrow OAM spectra because of its large Fermi momentum, but it can serve as a benchmark for deliberately broad distributions. Beyond postselection of a single OAM channel, future JINR experiments could therefore also investigate electron states with a controlled finite OAM dispersion and, with correlation-sensitive diagnostics, test whether such multimode states can be used for generating OAM-entangled states.}

\section{acknowledgements}

{We are especially grateful to Prof. P. Kazinski and V. Ryakin for their valuable discussions and clarifications concerning the theoretical model of the surface photoelectric effect. We also thank A. Dyatlov, I. Pavlov, A. Murtazin, and D. Kargina for fruitful discussions and helpful suggestions. The investigations of semiconductor characteristics are
supported by the Ministry of Science and Higher Education of the Russian Federation (Project FSER-2025-0012). The modeling of photoemission for accelerator photocathodes by A.~Shchepkin is supported by the Chinese Academy of Sciences President's International Fellowship Initiative (PIFI), Grant №~2026PVA0222. The calculations of the angular momentum transfer between photon and electron are supported by the Russian Science Foundation (Project №~23-62-10026; \url{https://rscf.ru/en/project/23-62-10026/}).}

\bibliographystyle{unsrtnat}
\bibliography{refs_formatted}
\end{document}